# Extragalactic electromagnetic cascades and cascade gamma-ray emission in magnetic fields of various strength

A.V. Uryson

Lebedev Physical Institute of the Russian Academy of Sciences Leninskii pr. 53 Moscow 119991

e-mail: uryson@sci.lebedev.ru

**Abstract**. We discuss the magnetic field influence on diffuse gamma-ray emission from extragalactic electromagnetic cascades initiated by ultra-high energy cosmic rays. Regions in space vary considerably in field strength: it is possibly of $10^{-12}$ G and lower in voids, of $\sim 10^{-6}$ G inside galaxies, galactic clusters and groups, of $\sim 10^{-7}$ G around them, and of $\sim 10^{-8}$-$10^{-9}$ G in filaments. Structures having fields higher than in voids occupy comparatively small fraction of the Universe, so they affect weakly on cascade emission. Still knowledge of this influence may be relevant studying large-scale component of the extragalactic magnetic field and to the search for exotic particles, as in the latter case contribution of all components to extragalactic gamma-ray background should be known, one of which is cascade emission. To study magnetic field effect we simulate particle propagation in homogeneous magnetic field of $\sim 10^{-6}$, $10^{-9}$, and $10^{-12}$ G and lower. It is found that in fields of $\sim 10^{-9}$ G and lower the spectra of diffuse cascade gamma-rays at energies $E \leq 10^{17}$ eV coincide. Thus no specific models of EGMF are required to study contribution of cascade emission in the extragalactic gamma-ray background at $E \leq 10^{17}$ eV. In the case of uniform field of $10^{-6}$ G (which seems to be unrealistic), this inference is valid in the energy range of $\sim 10^7$-$10^9$ eV. Results obtained can be also used studying large-scale component of the extragalactic magnetic field.

## 1. Introduction.

Soon after the cosmic microwave background (CMB) has been discovered it was shown that cosmic particles at ultra-high energies (UHE) $E>4\times 10^{19}$ eV interact with it loosing energy (GZK-effect) [Greisen, 1966; Zatsepin and Kuz'min, 1966] and giving rise to electromagnetic cascades [Hayakawa, 1966; Prilutsky and Rozental 1970]. UHECRs interact also with radio background emission [Berezinsky et al. 1990], and particles produced in cascades interact with CMB, radio emission and extragalactic background light (EBL).

Extragalactic cascades arise in the following processes. UHECRs assuming to be protons, interact in intergalactic space with CMB and radio emission mainly via reactions: $p+\gamma_r \rightarrow p+\pi^0$, $p+\gamma_r \rightarrow n+\pi^+$. Produced pions decay through the channel $\pi^0 \rightarrow \gamma+\gamma$, $\pi^+ \rightarrow \mu^+ + \nu_\mu$ giving rise to gamma-quanta, neutrinos and muons. Muons decay via $\mu^+ \rightarrow e^+ + \nu_e + \bar{\nu}_\mu$ and give rise to positrons and neutrinos. Neutrons decay producing p, $e^-$ and $\bar{\nu}_e$. Gamma-quanta, electrons and positrons are the particles which generate electromagnetic cascades in the interaction with background emission (CMB, radio, and EBL): $e+\gamma_b \rightarrow e' + \gamma'$ (IC scattering) and $\gamma+\gamma_b \rightarrow e^+ + e^-$ (pair production). Other reactions give a minor contribution to electromagnetic cascades. The development of the electromagnetic cascades in the universe in presence of CMB and EBL is described e.g. in [Berezinsky and Kalashev, 2016].

In addition to interaction with background emission, cascade electrons generate synchrotron radiation in the extragalactic magnetic field (EGMF), which quanta take part in IC scattering. The value of EGMF that weakly affects cascade process is estimated e.g. in [Uryson, 1998]: $B<10^{-9}$ G. Experimental indication of EGMF is analyzed intensively (see, e.g. [Kronberg, 1994, 2016; Beck 2009; Elyiv et al. 2009; Govoni et al. 2019; Abbasi et al. 2020; Dzhatdoev et al. 2020], but at present its structure and value has not yet clarified in detail. Diffuse cascade emission produced in homogeneous EGMF of various values was studied previously e.g. in [Beresinsky et al. 2011; Wang et al. 2011].

EGMF is apparently inhomogeneous, and regions in space vary considerably in field strength. Simulation of large-scale structure along with EGMF evolution was performed in e.g. [Dolag et al. 2005; Ryu et al. 1998; Ryu et al. 2008]. Recently codes ENZO and IllustrisTNG (TNG) has been applied to predict the evolution and present-day distribution of extragalactic magnetic fields: by Hackstein et al. (2016), Vazza et al. (2017) using a large suite of cosmological simulations with ENZO-code, and by Arámburo-García et al. (2021a, b) with a suite of large-volume cosmological simulations TNG.

Following results [Dolag et al. 2005; Ryu et al. 1998; Ryu et al. 2008; Vazza et al. 2017], magnetic field in voids is $B \leq 10^{-12}$ G (and has no effect on cascade process), it is of $\sim 10^{-6}$ G inside galaxies, galactic clusters and groups, of $\sim 10^{-7}$ G around them, and of $\sim 10^{-8} - 10^{-9}$ G in filaments. The limit of EGMF in voids approximately less than $10^{-11}$ G is derived analyzing CMB radiation in [Ade et al. 2016; Jedamzik and Saveliev, 2019]. Results by [Hackstein et al. 2016; Arámburo-García et al 2021a, b] demonstrate that space magnetization is influenced by processes of galaxy evolution, that might considerably enhance the fraction of space volume with $B>10^{-12}$ G, which might influence the propagation of UHE CRs.

In our paper we find out the EGMF influence on cascade emission, analyzing cases of homogeneous EGMF along particle way: a field of $10^{-12}$ G and lower, a field of $\sim 10^{-9}$ G, and of

~$10^{-6}$ G. This allowed us to avoid modelling of EGMF structure which is not reconstructed in detail.

Previously UHE CR cascade emission produced in homogeneous EGMF of values less than 1 nG was studied in [Berezinsky et al. 2011; Wang et al. 2011] (in order to obtain constraint on UHE CR emissivity satisfying data both on extragalactic gamma-ray background and cosmogenic neutrinos).

One of the first papers claiming that UHE CRs along with cascade emission can probe the large-scale component of the EGMF is [Lee et al. 1995] in which cascade gamma-ray spectra in EGMF of different values is analyzed. Lee et al. (1995) discuss gamma-ray energies above $10^{18}$ eV (at these energies cascade particle interaction with EBL are ignored). In our paper we also discuss the EGMF influence on the intensity of diffuse cascade gamma-quanta, but we analyze cascade emission in the energy range of $10^6 - 10^{20}$ eV. This range includes energies of gamma-rays available for Fermi-LAT covering the range of (20 MeV – 820 GeV) [Ackermann et al. 2015] along with planned gamma-ray telescope GAMMA-400 with the range of (1-500) GeV [Egorov et al. 2020], and future gamma-ray observatories CTA and SWGO with ranges (20 GeV-300 TeV) and (100s of GeV to PeV) respectively [Acharya et al. 2018; López-Coto et al. 2021]. Therefore our results can be used analyzing data of these gamma-ray telescopes and observatories.

In addition we choose the recent UHECR source model [Giacinti et al. 2015] which describes the bulk of CR data.

Our results could be of interest using both UHE CRs with their cascade emission as a probe for the EGMF, and when contribution of various components to the gamma-ray background should be known.

An example of the latter is hunting for dark matter particles, as it is expected that their annihilation or decay produce gamma-quanta in various objects (see e.g. [Roszkowski et al. 2018]. Then in order to filter a relatively weak hypothetical dark matter signal from the gamma-ray background the contribution of all components to the background should be known, including cascade gamma-rays.

The information about UHECR sources is needed to study electromagnetic cascades. This is CR sources, injection spectra, source distribution and evolution as sources can be located at distances according to red shifts up to 4-5.

UHECR sources have not been fully clarified. The general result was obtained by Hillas (1984): possible sites of particle acceleration should satisfy the condition between the size and magnetic field strength described by Eq.1 in [Hillas, 1984], as a result of which only objects above diagonal line in the plot "magnetic field strength-size" could accelerate protons to $10^{20}$ eV.

Source candidates satisfying this condition are active galactic nuclei (AGN) [Berezinsky et al. 1990], radio galaxy lobes ([Norman et al. 1995; Biermann, 1997] and ref. therein), and sites in SMBH vicinity. The latter case is analyzed in e.g. [Kardashev, 1995; Neronov et al. 2009].

UHECR injection spectra depend on processes of particle acceleration. CR accelerating by electric fields in the vicinity of SMBH or in jets, injection spectra are close to a monoenergetic one [Kardashev 1995; Neronov et al 2009; Istomin and Gunya 2020], while acceleration on shock fronts (existing e.g in AGN jets) results in exponential CR spectra: $\propto E^{-\alpha}$ with the spectral index $\alpha \approx 2.2$-$2.5$ [ Krymskii 1977; Bell 1978; Cesarsky 1992]. The unified model of CR acceleration and subsequent UHE proton propagation was suggested in [Giacinti et al. 2015] to describe data on fluxes of both CRs and diffuse gamma-rays and neutrinos. In this model sources have following parameters. First, a spectral slope of CR protons is $\alpha=2.2$. Second, the source red shift evolution is as that of Blue Lacertae objects (BL Lac), which are one of the AGN types. These source characteristics are used in our paper. For illustration we also consider another case of evolution that is history of cosmic star formation [Ÿuksel et al. 2008] in the form given by Wang et al. (2011). (Cosmic star formation rate possibly describes the evolution of UHE CR source density.)

Cascade gamma-ray intensity depends on UHE CR composition what has been studied in [Berezinsky et al. 2016] where pure proton composition or having admixture of Helium was analyzed. UHECR mass composition being a subject of active discussion, Hanlon et al. (2018) compare TA and PAO data and demonstrate that there is no contradiction between the data, thus UHECRs have mixed composition. Recently both TA and PAO [Abbasi et al. 2018; Aab et al. 2020] report on mixed composition including Helium and heavier nuclei up to Ferrum. Yet in our paper we study the EGMF influence on cascades from CR protons without analyses of cascade emission initiated by various nuclei.

UHECR propagation has been computed with the publically available code TransportCR [Kalashev and Kido, 2015], which is 1D code. It is valid because we calculate diffuse cascade emission initiated by randomly directed UHECRs. 3D simulation is applied when analyzing cascades which are originated by particles from a single source or in a blazar jet (see e.g. [Neronov and Aharonian 2007; Khalikov and Dzhatdoev, 2021]) or analyzing CR angle anisotropy (e.g. [Hackstein et al. 2018].

In our paper we find out the EGMF influence on cascade emission, analyzing cases of homogeneous EGMF along particle way: a field of $10^{-12}$ G and lower which is typical in the extragalactic space beyond clusters, sheets, and filaments, a field of $\sim 10^{-9}$ G which possibly fills sheets, and filaments, and of $\sim 10^{-6}$ G, which is common inside galaxies and galactic clusters. This allowed us to avoid modelling of EGMF structure which is not reconstructed in detail.

We obtain that in the EGMF with $B=10^{-9}$ G and $B \leq 10^{-12}$ G diffuse cascade emission is similar at energies $E \leq 10^{17}$ eV. Thus no specific models of EGMF are required to study contribution of cascade emission in the extragalactic gamma-ray background at these energies.

Considering EGMF of $10^{-6}$ G the result obtained is valid in the energy range of ~$10^7$-$10^9$ eV.

In the paper we use "photons" for particles of background emission, "quanta" for those produced in interaction with cosmic background and synchrotron process, and "electrons" for both electrons and positrons.

2. Method.

In this section we list assumptions underlying the model.

CRs are accelerated in processes in the vicinity of SMBH in galactic centers. As particle acceleration is connected with SMBH, AGNs can be UHECR sources regardless of their type and distance [Istomin and Gunya, 2020; Uryson, 2001].

We suppose that CRs are accelerated on shock fronts in vicinity of SMBH (e.g. in jets) which mechanism produces exponential injection spectra $\propto E^{-\alpha}$. In the model $\alpha=2.2$ following [Giacinti et al. 2015].

Distances from sources correspond to red shifts $z \approx 0.0$-5. The SMBH evolution is unclear. We use the evolution of Blue Lacertae objects (BL Lac), which are one of the AGN types, because the bulk of CR data is described with it [Giacinti et al. 2015]. Also we consider for illustration one more case of evolution when source density is proportional to the cosmic star formation rate [Ÿuksel et al. 2008]. It was used in the form given in [Wang et al. 2011].

UHECR mass composition evidently is mixed, which is discussed in Introduction. Yet in the model we assume for simplicity that UHECRs consist of protons.

Extragalactic background emissions are considered in the following way. The CMB has Planck energy distribution with the mean value $\varepsilon_r=6.7 \times 10^{-4}$ eV. The mean photon density is $n_r=400$ cm$^{-3}$. The background radio emission has parameters from the model of the luminosity evolution for radio galaxies [Protheroe and Biermann, 1996; 1997]. The EBL parameters are taken from [Inoue et al. 2013].

We assume the EGMF to be uniform and consider the field values $B=10^{-6}$, $10^{-9}$, and $B \leq 10^{-12}$ G.

3. Results and discussion.

Calculated gamma-ray spectra near the Earth in the fields of $10^{-6}$, $10^{-9}$, $10^{-12}$ G with two CR source evolution scenarios are shown in Fig. 1, 2. The spectra when $B<10^{-12}$ G coincide with that for $B=10^{-12}$ G, and are not shown in the figures.

In figures curves corresponding to same values of magnetic field are similar in shape, so evolution models under consideration have no effect on curve behavior.

At the energies $E \geq 10^{19}$ eV the calculated curves almost coincide as the bulk of the emission in this range is produced via $\pi^0$-decay, which is not affected by magnetic field. At lower energies quanta are produced both in cascades and in electron synchrotron emission. Synchrotron quanta being scattered by cascade particles, the energy of the latter transfers to a lower range, and a dip is formed on the curves in the range of $\sim 10^{14}$-$10^{18}$ eV. Larger is the field strength, the more is synchrotron emission and the more pronounced is the dip.

At energies $E \leq 10^{17}$ eV curves with $B=10^{-9}$ and $B=10^{-12}$ G coincide. This is in agreement with the result obtained for the range $E = 10^8$-$10^{14}$ eV in [Berezinsky et al. 2011].

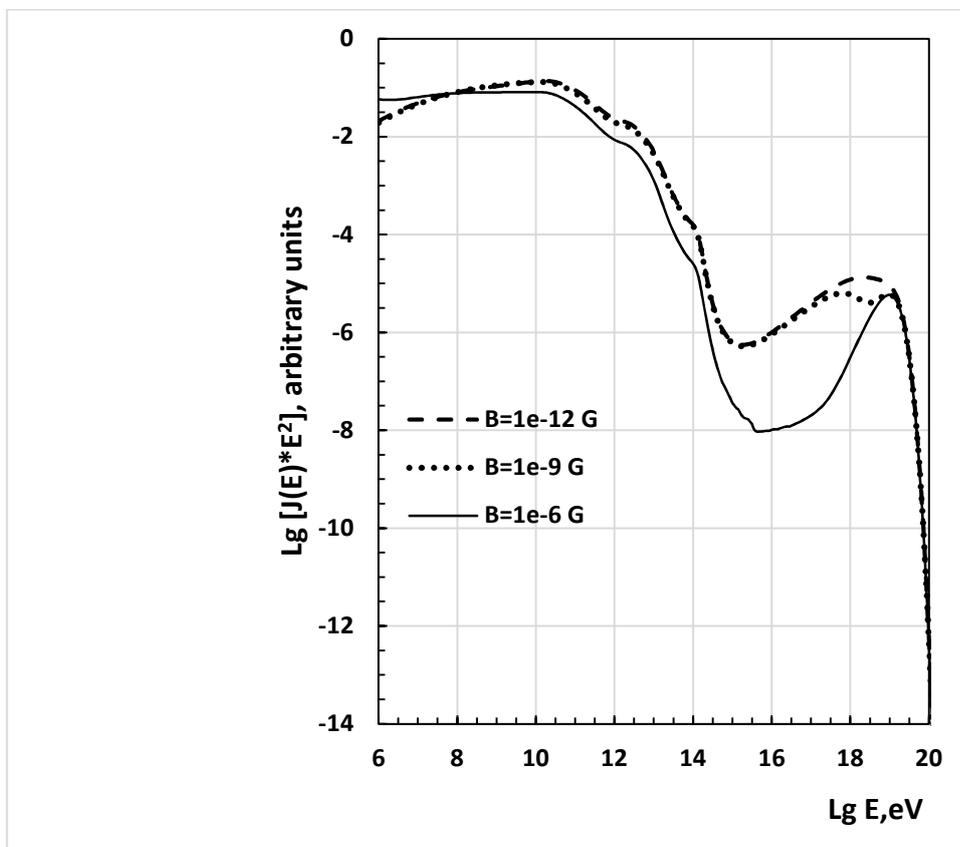

Fig.1. The spectra near the Earth of diffuse cascade gamma-rays produced by UHE protons. The evolution of UHECR sources is that of BL Lacs from [Giacinti et al. 2015]. The extragalactic magnetic field $B$ is of $10^{-6}$, $10^{-9}$ and $10^{-12}$ G. The spectra when EGMF is $B<10^{-12}$ G and $B=10^{-12}$ G coincide and are not shown in the figure. The maximal proton energy is of $4 \times 10^{20}$ eV.

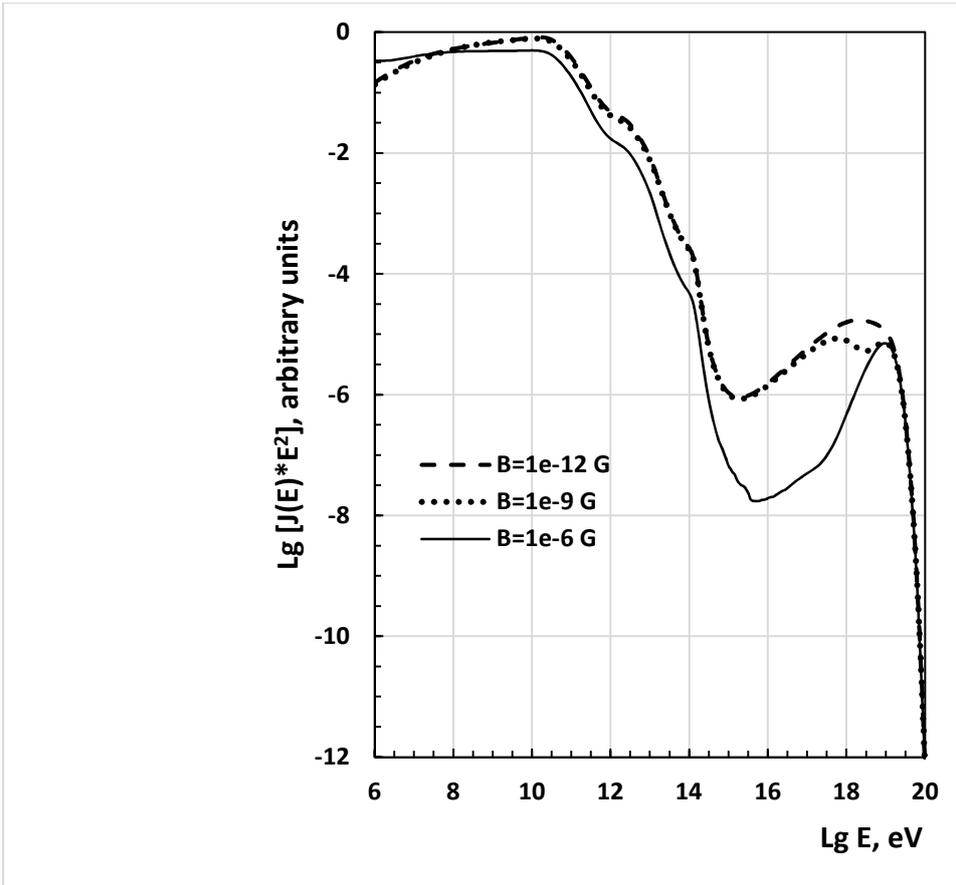

Fig.2. The same as in Fig.1, but UHE protons are from sources with evolution described by the star formation history [Ÿuksel et al. 2008] in the form [Wang et al. 2011].

Figures 1, 2 show that source evolution scenarios have no perceptible effect on the curve behavior. In contrast to this the calculated intensity of diffuse cascade gamma-ray emission depends on source evolution. This effect has been studied previously in e.g. [Kalashev et al. 2009; Wang et al. 2011; Berezinsky et al. 2016; Gavish and Eichler, 2016] (in which papers it was shown that source evolution affects on diffuse gamma-ray flux to UHECR flux ratio).

In both figures the curve for $B=10^{-6}$ G is different from the others both in shape and intensity. However in the energy range of ~$10^7$-$10^9$ eV the intensity when $B=10^{-6}$ is close to the others. This is illustrated in Fig. 3 where the ratio $R$ of cascade gamma-ray intensities in the range of $10^6$-$10^{11}$ eV in EGMF of $10^{-12}$ and $10^{-6}$ G is shown.

Equal intensities giving $R=1$, the relative deviation from unity is $\delta R \approx 0.25$-$0.3$ in the range of ~$10^7$-$10^9$ eV. At higher energies (to ~$10^{19}$ eV) the deviation is of ~10-100.

Based on the coincidence of curves for $B=10^{-9}$ and $B=10^{-12}$ G we conclude that no specific models of EGMF are required to study contribution of cascade emission in the extragalactic gamma-ray background at $E \leq 10^{17}$ eV. In the seemingly unrealistic case of EGMF of $10^{-6}$ G this inference is valid in the energy range of ~$10^7$-$10^9$ eV. (It is unrealistic because such field possibly fills only galaxies and their vicinity).

This result is obtained in the model where UHE CRs are protons. However CR nuclei possibly constitute a significant part of the UHE CR flux having a noticeable effect on the diffuse cascade emission (see e.g. [Berezinsky et al. 2016]). Study of UHE nuclei propagation and the resultant cascade gamma-ray emission will be performed in subsequent papers.

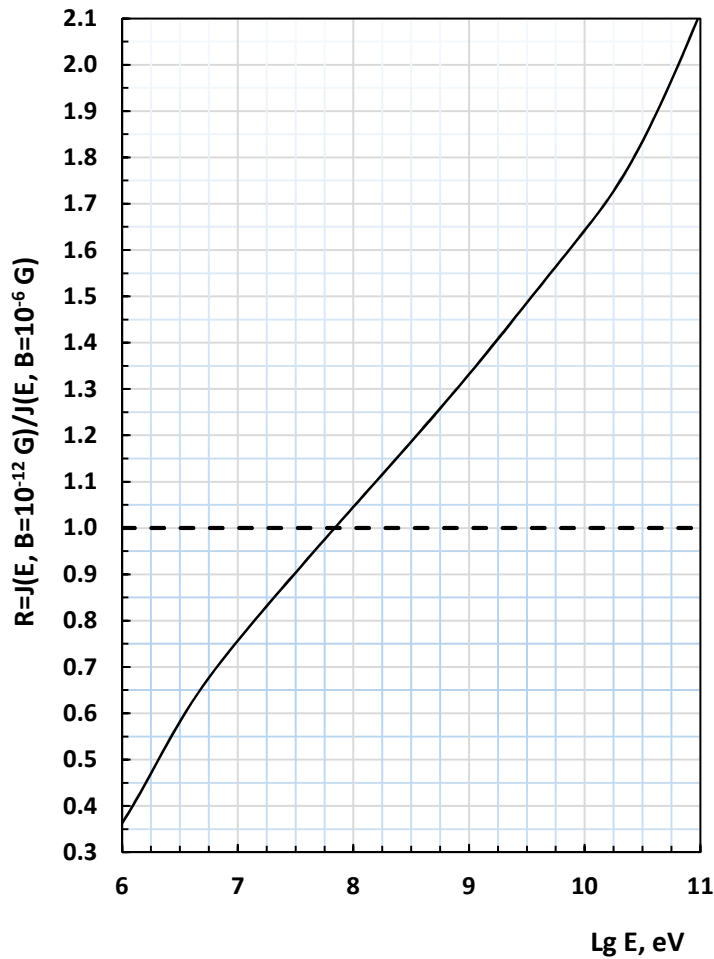

Fig. 3. The ratio $R$ of diffuse cascade gamma-ray intensities produced by UHE protons in the EGMF of $10^{-6}$ and $10^{-12}$ G in the energy range of $10^6$-$10^{11}$ eV. The horizontal dashed line shows $R=1$.

### 4. Conclusion.

Diffuse cascade gamma-ray emission depends on EGMF due to electron synchrotron radiation and IC scattering of its quanta.

We analyze the EGMF influence on the diffuse cascade gamma-ray spectra in the model with the uniform extragalactic magnetic field considering the extreme case of $10^{-6}$ G and more realistic cases of $10^{-9}$ and $10^{-12}$ G. The former is seemingly common inside galaxies and galactic clusters, and the two others are typical in the extragalactic space beyond clusters in sheets and filaments, and in voids. It is shown that cascade gamma-ray intensity in EGMF with $B = 10^{-9}$ and $B \leq 10^{-12}$ G are similar in the energy range $E \leq 10^{17}$ eV. Thus in this range specific models of extragalactic magnetic field are not required to study contribution of diffuse cascade emission in the extragalactic gamma-ray background. The result is obtained considering two possible scenarios of UHE CR source evolution. In the case of uniform EGMF of $10^{-6}$ G, which seems to be unrealistic, this inference is valid in the energy range of $\sim 10^7$-$10^9$ eV.

This result is seemingly valid also for purely electromagnetic cascades, as was discussed in [Dzhatdoev et al. 2019].

Structures where $B > 10^{-12}$ G fill the insignificant part of the space and their influence on the cascade emission is minor. Still our result could be of interest discussing UHE CRs and their cascade emission as a probe for the EGMF. The result obtained can be also relevant to the search for exotic particles, which are expected to annihilate or decay producing gamma-rays. Then in order to separate hypothetical dark matter signal from the gamma-ray background the contribution of all components to the background emission should be known, including cascade gamma-quanta.

We have considered the model where UHE CRs are protons. Study of UHE nuclei propagation and the resultant cascade gamma-ray emission will be performed in subsequent papers.

**Acknowledgements.**

The author thanks O. Kalashev for the discussion of the code TransportCR and extragalactic cascade features, E. Bugaev, T. Dzhatdoev, and M. Zelnikov for the discussion of extragalactic magnetic fields, A. Egorov for the discussion of dark matter signal registration. The author is grateful to referees for interesting discussion and remarks.

**References**

A. Aab, P. Abreu, M. Aglietta et al. Features of the energy spectrum of cosmic rays above $2.5 \times 10^{18}$ eV using the Pierre Auger Observatory. Phys. Rev. Lett. 125 121106 2020

R.U. Abbasi, M. Abe, T. Abu-Zayyad et al. Depth of ultra-high energy cosmic ray induced air shower maxima measured by the Telescope Array Black Rock and Long Ridge FADC fluorescence detectors and Surface Array in hybrid mode. Astrophys. J. 858 76 2018

R.U. Abbasi, M. Abe, T. Abu-Zayyad et al. Structure of Magnetic Deflection Multiplets of Ultra-High Energy Cosmic Rays. Astrophys. J. 899 86 2020

B.S. Acharya, I. Agudo, I. Al Samarai et al. Science with the Cherenkov Telescope Array. arXiv:1709.07997v2 [astro-ph.IM] 22 Jan 2018

M. Ackermann, M. Ajello, A. Albert et al. Limits on Dark Matter Annihilation Signals from the Fermi LAT 4-year Measurement of the Isotropic Gamma-Ray Background. JCAP 09 008 2015

P. A. R. Ade, N. Aghanim, M. Arnaud et al. Planck 2015 results. XIX. Constraints on primordial magnetic fields. Astron. Astrophys. A19 2016

A. Arámburo-García, K. Bondarenko, A. Boyarsky, D. Nelson, A. Pillepich, and A. Sokolenko. Magnetization of the intergalactic medium in the IllustrisTNG simulations: the importance of extended, outflow-driven bubbles. MNRAS 505 5038 2021a

A. Arámburo-García, K. Bondarenko, A. Boyarsky, D. Nelson, A. Pillepich, and A. Sokolenko. Ultra-high energy cosmic rays deflection by the intergalactic magnetic field. Phys. Rev. D 104, 083017 2021b

R. Beck. Galactic and extragalactic magnetic fields – a concise review. Astrophys. Space Sci. Trans. 5 43–47 2009

A.R. Bell. The acceleration of cosmic rays in shock fronts – I. MNRAS 182 147 1978

V. S. Berezinsky, S. V. Bulanov, V. A. Dogiel, V. L. Ginzburg and V. S. Ptuskin. *Astrophysics of cosmic rays,* Amsterdam, North-Holland, edited by V. L. Ginzburg. 1990

V. Berezinsky, A. Gazizov, M. Kachelrieß, S. Ostapchenko. Restricting UHECRs and cosmogenic neutrinos with Fermi-LAT. Physics Letters B 695 13 2011

V. Berezinsky, A. Gazizov, and O. Kalashev. Cascade photons as test of protons in UHECR. Astropart. Phys. 84 52 2016

V. Berezinsky and O. Kalashev. High energy electromagnetic cascades in extragalactic space: physics and features. Phys. Rev. D 94, 023007 2016

P. L. Biermann. The origin of the highest energy cosmic rays. J. *Phys. G: Nucl. Part. Phys.* 23 1 1997

C. J. Cesarsky. Cosmic rays with E > $10^{19}$ eV: Origin and transport. Nucl. Phys. B (Proc. Suppl.) 28B 51 1992

K. Dolag, D. Grasso, V. Springel and I. Tkachev. Constrained simulations of the magnetic field in the local Universe and the propagation of ultrahigh energy cosmic rays. JCAP 01 009 2005

T. Dzhatdoev, E. Khalikov, E. Podlesnyi, and A. Telegina. Intergalactic γ-ray propagation: basic ideas, processes, and constraints. Journal of Physics: Conf. Series 1181 012049 2019

T. A. Dzhatdoev, E. I. Podlesnyi and I. A. Vaiman. Can we constrain the extragalactic magnetic field from very high energy observations of GRB 190114C? Phys. Rev. D 102 123017 2020

A. E. Egorov, N. P. Topchiev, A. M. Galper et al. Dark matter searches by the planned gamma-ray telescope GAMMA-400. JCAP 11 049 2020

A. Elyiv, A. Neronov and D. V. Semikoz. Gamma-ray induced cascades and magnetic fields in the intergalactic medium. Phys. Rev. D 80 023010 2009

E. Gavish and D. Eichler. On ultra-high-energy cosmic rays and their resultant gamma-rays. Astrophys. J. 822 56 2016

G. Giacinti, M. Kachelrieß, O. Kalashev, A. Neronov, and D.V. Semikoz. Unified model for cosmic rays above $10^{17}$ eV and the diffuse gamma-ray and neutrino backgrounds. Phys. Rev. D 92 083016 2015

F. Govoni, E. Orrù, A. Bonafede et al. A radio ridge connecting two galaxy clusters in a filament of the cosmic web. Science 364 981 2019

K. Greisen. End to the Cosmic-Ray Spectrum? Phys. Rev. Lett. 16 748 1966

S. Hackstein, F. Vazza, M. Bruggen, G. Sigl, A. Dundovic. Propagation of Ultra High Energy Cosmic Rays in Extragalactic Magnetic Fields: A view from cosmological simulations. MNRAS 462 3660 2016

S. Hackstein, F. Vazza , M. Bruggen , J. G. Sorce , S. Gottlober. Simulations of ultra-high energy cosmic rays in the local universe and the origin of cosmic magnetic fields. MNRAS 475 2519 2018.

W. Hanlon, J. Bellido, J. Belz et al. Report of the working group on the mass composition of ultrahigh energy cosmic rays. *Proc.2016 Int. Conf. Ultra-High Energy Cosmic Rays (UHECR2016)* JPS Conf. Proc. 011013 2018

S. Hayakawa. Electron-Photon Cascade Process in Intergalactic Space. Prog. Theor. Phys. **37** 594 1966

A. M. Hillas. The origin of ultra-high energy cosmic rays. Ann. Rev. Astron. Astrophys. 22 425 1984

Y. Inoue, S. Inoue, M. Kobayashi et al. Extragalactic background light from hierarchical galaxy formation: gamma-ray attenuation up to the epoch of cosmic reionization and the first stars. Astrophys. J. **768** 197 2013

Ya. N. Istomin and A. A. Gunya. Acceleration of high energy protons in AGN relativistic jets. Phys. Rev. D 102 043010 2020

K. Jedamzik and A. Saveliev. Stringent limit on primordial magnetic fields from the cosmic microwave background radiation. Phys. Rev. Lett. 123, 021301 2019

O.E. Kalashev, D.V.Semikoz, and G.Sigl. Ultra-High Energy Cosmic Rays and the GeV-TeV Diffuse Gamma-Ray Flux. Phys. Rev. D 79 063005 2009 [arXiv:0704.2463 [astro-ph]].

O. E. Kalashev and E. Kido. Simulations of ultra-high-energy cosmic rays propagation. JETP 120 790 2015

N. S. Kardashev. Cosmic supercollider. MNRAS 276 515-520 1995

E. V. Khalikov and T. A. Dzhatdoev. Observable spectral and angular distributions of $\gamma$-rays from extragalactic ultrahigh energy cosmic ray accelerators: the case of extreme TeV blazars. MNRAS 505 1940 2021

P. P. Kronberg. *Cosmic magnetic fields*. Vol. 53 of Cambridge astrophysics series. Cambridge University Press 2016

P. P. Kronberg. Extragalactic magnetic fields. Rep. Progr. Phys. 325 382 1994

G. F. Krymskii. A regular mechanism for the acceleration of charged particles on the front of a shock wave. Akademiia Nauk SSSR Doklady. 234 1306 [Soviet Physics –Doklady **22** 327] 1977.

S. Lee, A. V. Olinto, and G. Sigl. Extragalactic magnetic field and the highest energy cosmic rays. Astrophys. J. Lett. 455 L21 1995 [arXiv: 9508088 [astro-ph]].

R. López-Coto, M. Doro, A. de Angelis, M. Mariotti, J. P. Harding. Prospects for the observation of primordial black hole evaporation with the southern wide field of view gamma-ray observatory. JCAP 08 40 2021

A. Neronov and F. A. Aharonian. Production of Tev gamma radiation in the vicinity of the supermassive black hole in the giant radio galaxy M87. Astrophys. J. 671 85-96 2007

A. Yu. Neronov, D. V. Semikoz and I. I. Tkachev. Ultra-high energy cosmic ray production in the polar cap regions of black hole magnetospheres. New Journal of Physics 11 065015 2009

C.A. Norman, D.B. Melrose, A. Achterberg. The origin of cosmic rays above $10^{18.5}$ eV. Astrophys. J. 454. 60 pp. 1995

O. Prilutsky and I.L. Rozental. Cascade processes in the Metagalaxy. Acta Phys. Hung. Suppl. 129 51 1970

R. J. Protheroe and P. L. Biermann. A new estimate of the extragalactic radio background and implications for ultra-high-energy gamma-ray propagation. Astropart. Phys. 6 45 1996; Erratum-ibid.7 181 1997

L. Roszkowski, E. M. Sessolo, and S. Trojanowski. WIMP dark matter candidates and searches – current status and future prospects. Rept.Prog.Phys. 81 06620 2018

D. Ryu, H. Kang, and P. L. Biermann. Cosmic magnetic fields in large scale filaments and sheets.

Astron. Astrophys. 335 19 1998

D. Ryu, H. Kang, J. Cho, and S. Das. Turbulence and magnetic fields in the large scale structure of the Universe. Science 320 909-912 2008

A. V. Uryson. Identification of extragalactic cosmic-ray sources using data from various detection facilities. JETP 89 597 1999

A. V. Uryson. Identification of sources of ultrahigh energy cosmic rays. Astronomy Reports 45 591 2001

F. Vazza, M. Bruggen , C. Gheller, S. Hackstein, D. Wittor, P. M. Hinz. Simulations of extragalactic magnetic fields and of their observables. *Class. Quantum Grav.* 34 234001 2017

X.-Y. Wang, R.-Y. Liu, and F. Aharonian. Constraining the emissivity of ultrahigh energy cosmic rays in the distant universe with the diffuse gamma-ray emission. Astrophys. J. 736 112 2011

H. Ÿuksel, M. D. Kistler, J. F. Beacom, and A. M. Hopkins. Revealing the high-redshift star formation rate with gamma-ray bursts. Astrophys. J. 683 L5 2008

G.T. Zatsepin and V.A. Kuz'min. Upper Limit of the Spectrum of Cosmic Rays. JETP Lett. 4 78 1966